\documentclass[10pt,journal]{IEEEtran}
\usepackage[english]{babel}
\usepackage[T1]{fontenc}
\usepackage[utf8]{inputenc}
\usepackage{float}
\usepackage{amsmath}
\usepackage{amssymb}
\usepackage{amsthm}
\usepackage{graphicx}
\usepackage{epstopdf}
\usepackage{subfigure}
\usepackage{babel}
\usepackage[keeplastbox]{flushend} %Note: This option has been added in version 3.0, released 2015/04/08, which can not be applied in the current Latex, but can be used correctly in arXiv,来解决arXiv在线编译中参考文献最后一行左定格的问题
\usepackage{indentfirst}
\usepackage{ifpdf}
\usepackage{cite}
\usepackage{stfloats}%h:here,t:top,b:buttom,p:page
\usepackage{cite}
\usepackage{bm}
\usepackage{mathtools}

\providecommand{\tabularnewline}{\\}
\floatstyle{ruled}
\newfloat{algorithm}{tbp}{loa}
\floatname{algorithm}{\protect\algorithmname}

\IEEEoverridecommandlockouts

\ifCLASSINFOpdf
\else
\fi

\newtheorem{mypro}{\hspace{1em}\it Proposition}

\newtheorem{remark}{\hspace{1em}\it Remark}
\newenvironment{myproof}{\it Proof:}{\hfill $\blacksquare$\par}

\makeatletter
\g@addto@macro{\thm@space@setup}{\thm@headpunct{:}} % set the format of remark
\renewcommand{\@upn}{} % to use the same font for the number as for the head
\makeatother

\begin{document}
%\selectlanguage{ENGLISH}
\renewcommand{\figurename}{Fig.}

\title{Low-Complexity Linear Equalizers for OTFS Exploiting Two-Dimensional Fast Fourier Transform}

\author{Junqiang~Cheng,
        Hui~Gao, \IEEEmembership{Senior Member,~IEEE,}
        Wenjun~Xu, \IEEEmembership{Senior Member,~IEEE,}\\
        Zhisong~Bie, and Yueming Lu% <-this % stops a space
\thanks{The authors are with the School
of Information and Communication Engineering, Beijing University of Posts and Telecommunications, Beijing, 100876, China (e-mail: jqcheng@bupt.edu.cn; huigao@bupt.edu.cn).}}
\maketitle
\begin{abstract}
Orthogonal time frequency space (OTFS) modulation can effectively convert a doubly dispersive channel into an almost non-fading channel in the delay-Doppler domain. However, one critical issue for OTFS is the very high complexity of equalizers. In this letter, we first reveal the doubly block circulant feature of OTFS channel represented in the delay-Doppler domain. By exploiting this unique feature, we further propose zero-forcing (ZF) and minimum mean squared error (MMSE) equalizers that can be efficiently implemented with the two-dimensional fast Fourier transform. The complexity of our proposed equalizers is gracefully reduced from $\mathcal{O}\left(\left(NM\right)^{3}\right)$ to $\mathcal{O}\left(NM\mathrm{log_{2}}\left(NM\right)\right)$, where $N$ and $M$ are the number of OTFS symbols and subcarriers, respectively. Analysis and simulation results show that compared with other existing linear equalizers for OTFS, our proposed linear equalizers enjoy a much lower computational complexity without any performance loss.
\end{abstract}


\section{Introduction}
\IEEEPARstart{H}{igh}-mobility scenario is one of the key scenarios of future broadband wireless communication networks. However, the time-variant channel in such scenario causes significant inter-carrier interference (ICI), which seriously degrades the performance of orthogonal frequency division multiplexing (OFDM) systems \cite{wang2006performance}. Orthogonal time frequency space (OTFS) modulation, which was originally proposed in \cite{hadani2017otfs}, has been demonstrated to achieve a significant performance improvement over OFDM in doubly dispersive channels. A key hallmark of OTFS is the conversion of a doubly dispersive channel into an almost non-fading channel in the delay-Doppler domain, and consequently all symbols in a frame experience a relatively stable channel \cite{monk2016otfs}.

The complexity of equalization is critical to practical OTFS systems. There are some recent works on the equalizer design for OTFS, where the most studied one is the message passing (MP) scheme \cite{raviteja2018interference}. However, the complexity of this non-linear equalization scheme depends on the sparsity level of the channel in the delay-Doppler domain, which might not be guaranteed in the realistic scenarios. On the contrary, linear equalization usually enjoys a much simpler structure with lower complexity, and are thus more suitable for practical applications. Moreover, linear equalization is often used as a pre-processing in some nonlinear schemes, e.g., decision feedback equalizer \cite{ding2019otfs}. Therefore, in this letter, we focus on the design of low-complexity linear equalizers for OTFS.

 As for the typical linear equalizers, such as zero-forcing (ZF) and minimum mean squared error (MMSE) equalizers, the operation of matrix inversion is often inevitable \cite{hadani2017otfs}, which results in a prohibitively high computational complexity of $\mathcal{O}\left(\left(NM\right)^{3}\right)$, where $N$ and $M$ are the number of OTFS symbols and subcarriers, and both are very large in OTFS systems. Noting this limitation, a frequency-domain ZF (FD-ZF) equalizer is investigated in \cite{ding2019otfs}, which utilizes the Kronecker product based on the block circulant feature of the channel to avoid the matrix inversion. However, it still suffers a computational complexity up to $\mathcal{O}\left(\left(NM\right)^{2}\right)$.

In this letter, we first analyze the doubly block circulant (DBC) feature of the OTFS channel represented in the delay-Doppler domain. More specifically, the channel matrix is carefully arranged in a block circulant matrix form, wherein each submatrix is a circulant matrix. Then, we show that the DBC channel matrix can be diagonalized by two-dimensional (2D) discrete Fourier transform (DFT) matrix and 2D inverse DFT (IDFT) matrix, which inspires us to obtain the output of the ZF and MMSE equalizers by the 2D fast Fourier transform (FFT) operation without conventional full matrix inversion. As a beneficial result, the complexity of our proposed linear equalizers is significantly reduced to $\mathcal{O}\left(NM\mathrm{log_{2}}\left(NM\right)\right)$. Finally, simulation results demonstrate that compared with other existing linear equalizers for OTFS, our proposed linear equalizers enjoy a much lower computational complexity without any performance loss.\footnote{\emph{Notation:} Matrix, vector and scalar are denoted by $\mathbf{A}$, $\mathbf{a}$ and $a$, respectively.  The function $\mathrm{vec}\left(\mathbf{A}\right)$ denotes the columns-wise matrix vectorization, and its inverse operation is $\mathrm{unvec}\left(\mathbf{a}\right)$. The function $\mathrm{diag\left(\mathbf{a}\right)}$ forms a diagonal matrix with the elements of $\mathbf{a}$. The superscripts $\left(\cdot\right)^{T}$ and $\left(\cdot\right)^{H}$ indicate transpose and conjugate transpose operations, respectively. Finally, the operators $\otimes$ and $\left\langle \cdot\right\rangle _{N}$ represent Kronecker product and modulo-$N$ operations, respectively.}

\section{OTFS System Model}

\begin{figure*}
  \centering
  \includegraphics[scale=0.6]{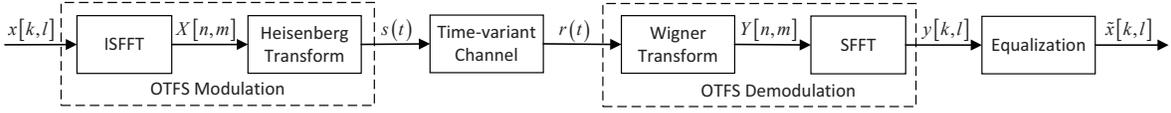}
  \caption{OTFS system model.}
  \label{fg_otfs}
\end{figure*}

In this section, we briefly review the OTFS system model, and formulate the input-output relation in a vectorized form.

\subsection{Channel model in the delay-Doppler domain}
A sparse representation of the wireless channel in the delay-Doppler domain can be expressed as \cite{raviteja2018low}
\begin{align}
h\left(\tau,\nu\right)=\sum_{i=1}^{P}h_{i}\delta\left(\tau-\tau_{i}\right)\delta\left(\nu-\nu_{i}\right),
\end{align}
where $P$ is the number of propagation paths, $h_{i}$, $\tau_{i}$ and $\nu_{i}$ denote the channel gain, delay and Doppler shift associated with the $i$-th propagation path, respectively. Within the OTFS framework, the discretized time-frequency (TF) plane is obtained by sampling time and frequency axes at intervals $T$ (seconds) and $\Delta f$ (Hz), respectively,
\begin{equation*}
\Lambda=\left\{ \left(nT,m\Delta f\right),n=0,\cdots,N-1,m=0,\cdots,M-1\right\}.
\end{equation*}
The corresponding discretized delay-Doppler plane is
\begin{equation*}
\Gamma\!=\!\left\{\! \left(\frac{k}{NT},\frac{l}{M\Delta f}\right)\!,k=0,\cdots\!,N-1,l=0,\cdots\!,M-1\right\}\!,
\end{equation*}
where $\frac{1}{NT}$ and $\frac{1}{M\Delta f}$ represent the resolution of Doppler and delay axes, respectively.

Based on the above notations, the delay and Doppler taps for the $i$-th path of $h\left(\tau,\nu\right)$ are given as \cite{ding2019otfs}
\begin{align}
\tau_{i}=\frac{l_{\tau_{i}}+\tilde{l}_{\tau_{i}}}{M\Delta f},\: \nu_{i}=\frac{k_{\nu_{i}}+\tilde{k}_{\nu_{i}}}{NT},
\end{align}
where  integers $l_{\tau_{i}}$ and $k_{\nu_{i}}$ denote the indexes of the delay tap and Doppler tap, respectively. While  $\tilde{l}_{\tau_{i}}$ and $\tilde{k}_{\nu_{i}}$ correspond to the fractional delay and the fractional Doppler shift, respectively. Next, we formulate the input-output relation of OTFS in a vectorized form with the help of the above definitions.

\subsection{Orthogonal time frequency space modulation}
As shown in Fig. \ref{fg_otfs}, OTFS modulation first maps the 2D information symbols $x\left[k,l\right],k=0,\cdots,N-1,l=0,\cdots,M-1$ in the delay-Doppler domain into TF domain through \emph{inverse symplectic finite Fourier transform} (ISFFT),
\begin{align}
X\left[n,m\right]=\frac{1}{NM}\sum_{k=0}^{N-1}\sum_{l=0}^{M-1}x\left[k,l\right]e^{j2\pi\left(\frac{nk}{N}-\frac{ml}{M}\right)},
\end{align}
where $n=0,\cdots,N-1$, $m=0,\cdots,M-1$.

Next, the TF symbols $X\left[n,m\right]$ are transformed to the time domain signal $s\left(t\right)$ through Heisenberg transform,
\begin{align}
s\left(t\right)=\sum_{n=0}^{N-1}\sum_{m=0}^{M-1}X\left[n,m\right]g_{t}\left(t-nT\right)e^{j2\pi m\Delta f\left(t-nT\right)},
\end{align}
where $g_{t}\left(t\right)$ is the transmit pulse. Correspondingly, there is also a receive pulse $g_{r}\left(t\right)$ in Wigner transform, the inverse of Heisenberg transform, at the receiver. By assuming the ideal orthogonality between $g_{t}\left(t\right)$ and $g_{r}\left(t\right)$ \cite{hadani2017otfs}, i.e., the cross-ambiguity function of them is zero, the received signal in the TF domain can be formulated as
\begin{equation}
Y\left[n,m\right]=H\left[n,m\right]X\left[n,m\right]+Z\left[n,m\right],
\end{equation}
where $Z\left[n,m\right]$ is the white Gaussian noise in the TF domain, and
\begin{equation}
H\left[n,m\right]=\iint h\left(\tau,\nu\right)e^{j2\pi\nu nT}e^{-j2\pi\left(\nu+m\Delta f\right)\tau}d\nu d\tau.
\end{equation}

Finally, by applying \emph{symplectic finite Fourier transform} (SFFT) to the TF symbols $Y\left[n,m\right]$, we obtain the received 2D information symbols $y\left[k,l\right]$ in the delay-Doppler domain,
\begin{equation}
\begin{aligned}
&y\left[k,l\right]=\frac{1}{NM}\sum_{n=0}^{N-1}\sum_{m=0}^{M-1}Y\left[n,m\right]e^{-j2\pi\left(\frac{nk}{N}-\frac{ml}{M}\right)}\\
&=\frac{1}{NM}\sum_{n=0}^{N-1}\sum_{m=0}^{M-1}x\left[k,l\right]h_{w}\left[\frac{k-n}{NT},\frac{l-m}{M\Delta f}\right]+z\left[k,l\right],
\label{eq_y0}
\end{aligned}
\end{equation}
where $z\left[k,l\right]$ is the complex Gaussian noise in the delay-Doppler domain. $h_{w}\left[\cdot,\cdot\right]$ is a discrete version of $h_{w}\left(\nu{'},\tau{'}\right)$, which is the circular convolution of the channel response $h\left(\tau,\nu\right)$ with the ISFFT of a rectangular windowing function in the TF domain. Following \cite{ding2019otfs, raviteja2018low, surabhi2018diversity}, $N$ and $M$ are sufficiently large in practical systems, so that the fractional delay and Doppler shift can be ignored (i.e., $\tilde{l}_{\tau_{i}}=\tilde{k}_{\nu_{i}}=0$).

With the above assumptions, the input-output relation of OTFS system in (\ref{eq_y0}) can be formulated as \cite{raviteja2018low}
\begin{equation}
y\left[k,l\right]=\sum_{i=1}^{P}h_{i}^{'}x\left[\left\langle k-k_{\nu_{i}}\right\rangle _{N},\left\langle l-l_{\tau_{i}}\right\rangle _{M}\right]+z\left[k,l\right],
\label{eq_y}
\end{equation}
where $h_{i}^{'}=h_{i}e^{-j2\pi\nu_{i}\tau_{i}}$. The above equation can be further represented in a vectorized form as \cite{raviteja2018low}
\begin{equation}
\mathbf{y}=\mathbf{Hx}+\mathbf{z},
\label{eq_system}
\end{equation}
where $\mathbf{x},\mathbf{y},\mathbf{z}\in\mathbb{C}^{NM\times1}$, $\mathbf{H}\in\mathbb{C}^{NM\times NM}$. The $\left(k+Nl\right)$-th element of $\mathbf{y}$ is $\left[\mathbf{y}\right]_{k+Nl}=y\left[k,l\right],k=0,\cdots,N-1,l=0,\cdots,M-1$. The same relation holds for $\mathbf{x}$ and $\mathbf{z}$ as well. Next, we will propose our low-complexity linear equalizers based on this input-output model.

\section{Low-Complexity Linear Equalizers for OTFS }
In this section, we briefly review the classical ZF and MMSE equalizers, both of which are based on the matrix inversion operation. Then we analyse the DBC channel structure of OTFS. Based on it, we propose to obtain the output of the ZF and MMSE equalizers for OTFS by 2D FFT operations, and hence significantly reduce the computational complexity.

\subsection{Direct channel matrix inversion based linear equalizers}
According to the system model in (\ref{eq_system}), the output of the ZF equalizer with full matrix inversion can be expressed as
\begin{equation}
\mathbf{\tilde{x}}_{ZF}=\mathbf{W}_{ZF}\mathbf{y}=\left(\mathbf{H}^{H}\mathbf{H}\right)^{-1}\mathbf{H}^{H}\mathbf{y}.
\label{eq_c_zf}
\end{equation} In a similar fashion, the output of the MMSE equalizer that takes the noise variance $\sigma_{z}^{2}$ into account is expressed as
\begin{equation}
\mathbf{\tilde{x}}_{MMSE}=\mathbf{W}_{MMSE}\mathbf{y}=\left(\mathbf{H}^{{H}}\mathbf{H}+\sigma_{z}^{2}\mathbf{I}_{NM}\right)^{-1}\mathbf{H}^{{H}}\mathbf{y}.
\label{eq_c_mmse}
\end{equation}

Note that the full matrix inversion operations in (\ref{eq_c_zf}) and (\ref{eq_c_mmse}) require $\mathcal{O}\left(\left(NM\right)^{3}\right)$ complex multiplications \cite{schniter2004low}, which is prohibitively complex for large symbol length. In fact, the OTFS channel $\mathbf{H}$ represented in the delay-Doppler domain enjoys a DBC feature, which bears a potential to significantly reduce the computational complexity of the linear equalizers. Next, we will make a detailed analysis on this point.

\subsection{Doubly block circulant channel structure of OTFS}
Considering the transmitted information symbols with $M$ subcarriers and $N$ OTFS symbols, the ISFFT and SFFT operations at the transmitter and receiver, respectively, inherently lead to the 2D circular convolution in (\ref{eq_y}), which can be represented in a vector form as (\ref{eq_system}). Therefore, the channel matrix $\mathbf{H}$ in (\ref{eq_system}) is a DBC matrix with the structure
\begin{align}
\mathbf{H}=\left[\begin{array}{ccccc}
\mathbf{H}_{0} & \mathbf{H}_{M-1} & \cdots & \mathbf{H}_{2} & \mathbf{H}_{1}\\
\mathbf{H}_{1} & \mathbf{H}_{0} & \cdots & \mathbf{H_{\mathrm{3}}} & \mathbf{H}_{2}\\
\vdots & \vdots & \ddots & \vdots & \vdots\\
\mathbf{H}_{M-2} & \mathbf{H}_{M-3} & \cdots & \mathbf{H}_{0} & \mathbf{H}_{M-1}\\
\mathbf{H}_{M-1} & \mathbf{H}_{M-2} & \cdots & \mathbf{H}_{1} & \mathbf{H}_{0}
\label{eq_10}
\end{array}\right],
\end{align}
where each submatrix $\mathbf{H}_{m},m=0,\cdots,M-1$, is an $N\times N$ circulant matrix determined by (\ref{eq_y}).

% Let $\mathbf{H}^{'}=\mathrm{unvec}\left(\mathbf{h}\right)$, where $\mathbf{h}$ is the first column of $\mathbf{H}$. Obviously, the DBC matrix $\mathbf{H}$ is completely determined by $\mathbf{H^{'}}$ and the circulant structure.
%we will make use of the characterization of the eigenvalues and eigenvectors of the DBC matrices, which uses the following definition:
%\begin{equation}
%\mathbf{\Xi}\overset{def}{=}\mathbf{F}_{M}\otimes\mathbf{F}_{N},
%\end{equation}
%where $\mathbf{F}_{M}$ represents the normalized $M$-point DFT matrix. The matrix $\mathbf{\Xi}$ is the so-called 2D DFT matrix \cite{jain1989fundamentals}, and we also introduce the inverse 2D DFT matrix $\mathbf{\Xi}^{-1}=\left(\mathbf{F}_{M}\otimes\mathbf{F}_{N}\right)^{-1}$.

Next, we introduce the normalized $M$-point DFT matrix $\mathbf{F}_{M}$, the 2D DFT matrix \cite{jain1989fundamentals}
\begin{equation}
\mathbf{\Xi}\overset{\Delta}{=}\mathbf{F}_{M}\otimes\mathbf{F}_{N},
\end{equation}
and its inverse, $\mathbf{\Xi}^{-1}=\left(\mathbf{F}_{M}\otimes\mathbf{F}_{N}\right)^{-1}$, which will be used in the following proposition to decompose the DBC channel matrix $\mathbf{H}$.
\begin{mypro}
\emph{As a DBC matrix in the form of (\ref{eq_10}), the channel matrix $\mathbf{H}$ has the following eigen-decomposition
\begin{equation}
\mathbf{H}=\mathbf{\Xi}^{-1}\mathbf{\Delta}\mathbf{\Xi},
\label{eq_H}
\end{equation}
where
\begin{equation}
\mathbf{\Delta}=\mathrm{diag}\left(\mathrm{vec}\left(\mathbf{F}_{N}\mathbf{H}^{'}\mathbf{F}_{M}\right)\right),
\label{eq_deltaH}
\end{equation}
is a diagonal matrix, which is composed of the eigenvalues of $\mathbf{H}$. Besides, $\mathbf{H}^{'}=\mathrm{unvec}\left(\mathbf{h}_{1}\right)$ is made up by the first column of $\mathbf{H}$}.
\end{mypro}

\begin{myproof}
\emph{The columns of the 2D IDFT matrix $\mathbf{\Xi}^{-1}$ are the eigenvectors of any $NM\times NM$ DBC matrix \cite{jain1989fundamentals}. Therefore, the channel matrix $\mathbf{H}$ satisfies
\begin{equation}
\mathbf{H}\bm{\xi}_{q}=\delta_{q}\bm{\xi}_{q},
\end{equation}
where $\bm{\xi}_{q}$, $q=0,\cdots,NM-1$, is the $q$-th column of $\mathbf{\Xi}^{-1}$, and $\delta_{q}$ is the corresponding eigenvalue. The above formula can be further written in terms of $\mathbf{\Xi}^{-1}$,
\begin{equation}
\mathbf{H}\mathbf{\Xi}^{-1}=\mathbf{\Xi}^{-1}\mathbf{\Delta},
\end{equation}
which is equivalent to (\ref{eq_H}) in form with $\mathbf{\Delta}=\mathrm{diag}\left(\bm{\delta}\right)$ and $\bm{\delta}=\left[\delta_{0},\cdots,\delta_{NM-1}\right]$. Also note the eigenvalues of the DBC matrix $\mathbf{H}$ can be calculated by the 2D DFT of its first column $\mathbf{h}_{1}$ \cite{jain1989fundamentals}, i.e.,
\begin{equation}
\bm{\delta}=\mathbf{\Xi}\mathbf{h}_{1}=\left(\mathbf{F}_{M}\otimes\mathbf{F}_{N}\right)\mathbf{h}_{1}.
\label{eq_smalldelta}
\end{equation}
Using $\mathbf{F}_{M}^{T}=\mathbf{F}_{M}$ and $\mathrm{vec}\left(\mathbf{AXB}\right)=\left(\mathbf{B}^{T}\otimes\mathbf{A}\right)\mathrm{vec}\left(\mathbf{X}\right)$, we can rewrite (\ref{eq_smalldelta}) as
\begin{equation}
\bm{\delta}=\mathrm{vec}\left(\mathbf{F}_{N}\mathbf{H}^{'}\mathbf{F}_{M}\right),
\end{equation}
then the proof is finished.}
\end{myproof}

\begin{remark}
\emph{It is obvious that the diagonal matrix $\mathbf{\Delta}$ (i.e., the eigenvalues of $\mathbf{H}$) is completely determined by $\mathbf{H}^{'}$, as analyzed later, which in turn can be used to obtain the output of our proposed linear equalizers in a very efficient way.
}
\end{remark}

%\begin{mypro}
%\emph{The diagonal matrix $\mathbf{\Delta}$ can be obtained with only the first column of $\mathbf{H}$, i.e.,
%\begin{equation}
%\mathbf{\Delta}=\mathrm{diag}\left(\mathrm{vec}\left(\mathbf{F}_{N}\mathbf{H}^{'}\mathbf{F}_{M}\right)\right).
%\label{eq_deltaH}
%\end{equation}}
%\end{mypro}
%
%\begin{myproof}
%\emph{For any DBC matrix, its eigenvalues comprise the 2D DFT values of the first column of the DBC matrix \cite{jain1989fundamentals}. Therefore, we can get the following formula:
%\begin{equation}
%\bm{\delta}=\mathbf{\Xi}\mathbf{h},
%\label{eq_smalldelta}
%\end{equation}
%where $\bm{\delta}=\left[\delta_{0},\cdots,\delta_{NM-1}\right]$. Thanks to the properties $\mathbf{F}_{M}^{T}=\mathbf{F}_{M}$ and $\mathrm{vec}\left(\mathbf{AXB}\right)=\left(\mathbf{B}^{T}\otimes\mathbf{A}\right)\mathrm{vec}\left(\mathbf{X}\right)$, we can also rewrite (\ref{eq_smalldelta}) as
%\begin{equation}
%\bm{\delta}=\mathrm{vec}\left(\mathbf{F}_{N}\mathbf{H}^{'}\mathbf{F}_{M}\right).
%\end{equation}
%Therefore, the diagonal matrix $\mathbf{\Delta}$ can be obtained as
%\begin{equation}
%\mathbf{\Delta}=\mathrm{diag}\left(\bm{\delta}\right)=\mathrm{diag}\left(\mathrm{vec}\left(\mathbf{F}_{N}\mathbf{H}^{'}\mathbf{F}_{M}\right)\right).
%\end{equation}
%The proof of the proposition 2 is complete.}
%\end{myproof}

\subsection{2D FFT based ZF and MMSE equalizers for OTFS}
In contrast to the conventional ZF/MMSE equalizers in (\ref{eq_c_zf}) and (\ref{eq_c_mmse}) with complicated direct matrix inversion, in this subsection, we propose novel 2D FFT based ZF and MMSE equalizers for OTFS, which carefully exploit the DBC structure of the channel matrix and enable very efficient 2D FFT implementation without the complicated matrix inversion.

More specifically, note that FFT is an efficient implementation of DFT, we first introduce the following new operation
\begin{equation}
\begin{aligned}
\mathrm{vFFT_{2}}\left(\mathbf{A}\right)\overset{\Delta}{=}\mathrm{vec}\left(\mathbf{F}_{N}\mathbf{A}\mathbf{F}_{M}\right)=\mathbf{\Xi}\times\mathrm{vec}\left(\mathbf{A}\right),
\end{aligned}
\end{equation}
where $\mathbf{A}\in\mathbb{C}^{N\times M}$. The new operation mainly includes the 2D FFT over matrix $\mathbf{A}$, i.e., FFT to each of the $M$ columns of $\mathbf{A}$ and then to each of the $N$ rows of the resulting matrix \cite{brigham1988fast}, and finally the vectorization of its result. In a similar way, we can introduce $\mathrm{vIFFT_{2}}\left(\mathbf{A}\right)\overset{\Delta}{=}\mathbf{\Xi}^{-1}\times\mathrm{vec}\left(\mathbf{A}\right)$. Subsequently, we can effectively calculate (\ref{eq_deltaH}) as
\begin{equation}
\mathbf{\Delta}=\mathrm{diag}\left(\mathrm{vFFT_{2}}\left(\mathbf{H}^{'}\right)\right).
\end{equation}

Armed with these new 2D FFT operators and key results, we can continue to propose the 2D FFT based ZF ($\mathrm{FFT}_{2}$-ZF) and MMSE ($\mathrm{FFT}_{2}$-MMSE) equalizers as follows.
\subsubsection{$\mathrm{FFT}_{2}$-ZF equalizer}
First, with reference to the system model in (\ref{eq_system}) and Proposition 1, we can rewrite the received signal as
\begin{equation}
\mathbf{y}=\mathbf{\Xi}^{-1}\mathbf{\mathbf{\Delta}}\mathbf{\Xi}\mathbf{x}+\mathbf{z},
\label{eq_ynew}
\end{equation}
then the output of the $\mathrm{FFT}_{2}$-ZF equalizer can be obtained as
\begin{equation}
\begin{aligned}
\mathbf{x}_{ZF}=\mathbf{\Xi}^{-1}\mathbf{\mathbf{\Delta}}^{-1}\mathbf{\Xi}\mathbf{y}=\mathbf{\mathrm{vIFFT_{2}}}\left(\mathrm{unvec}\left(\mathbf{v}_{ZF}\right)\right),
\label{eq_ZF}
\end{aligned}
\end{equation}
where
\begin{equation}
\mathbf{v}_{ZF}=\mathbf{\Delta}^{-1}\times\mathrm{v}\mathrm{FFT}_{2}\left(\mathrm{\mathbf{\mathrm{unvec}}}\left(\mathbf{y}\right)\right).
\end{equation}

\subsubsection{$\mathrm{FFT}_{2}$-MMSE equalizer}
Similarly, we can formulate the output of the $\mathrm{FFT}_{2}$-MMSE equalizer as
\begin{equation}
\begin{aligned}
\mathbf{x}_{MMSE}&=\mathbf{\Xi}^{-1}\left(\mathbf{\Delta}^{{H}}\mathbf{\Delta}+\sigma_{z}^{2}\mathbf{I}_{NM}\right)^{-1}\mathbf{\Delta}^{{H}}\mathbf{\Xi}\mathbf{y}\\
&=\mathrm{vIFFT_{2}}\left(\mathrm{unvec}\left(\mathbf{v}_{MMSE}\right)\right),
\label{eq_MMSE}
\end{aligned}
\end{equation}
where
\begin{equation}
\mathbf{v}_{MMSE}=\left(\mathbf{\Delta}^{{H}}\mathbf{\Delta}+\sigma_{z}^{2}\mathbf{I}_{NM}\right)^{-1}\mathbf{\Delta}^{{H}}\times\mathrm{v}\mathrm{FFT}_{2}\left(\mathrm{\mathbf{\mathrm{unvec}}}\left(\mathbf{y}\right)\right).
\end{equation}

\renewcommand\arraystretch{1.7}
\begin{table}[htb]
\noindent
\centering
\caption{Computational Complexity Comparison of Different Linear Equalizers for OTFS}
\begin{tabular}{|c|c|c|}
\hline
\textbf{Equalizer}  & \textbf{Computational Complexity}\tabularnewline
\hline
Matrix inversion based ZF/MMSE \cite{hadani2017otfs} & $\mathcal{O}\left(\left(NM\right)^{3}\right)$\tabularnewline
\hline
FD-ZF utilizing Kronecker product \cite{ding2019otfs} & $\mathcal{O}\left(\left(NM\right)^{2}\right)$\tabularnewline
\hline
$\mathrm{FFT}_{2}$-ZF/MMSE & $\mathcal{O}\left(NM\mathrm{log_{2}}\left(NM\right)\right)$\tabularnewline
\hline
\end{tabular}
\label{tb}
\end{table}

\begin{remark}
\emph{It is remarkable that the most complex operation of our proposed $\mathrm{FFT}_{2}$-ZF and $\mathrm{FFT}_{2}$-MMSE equalizers is the 2D FFT/2D IFFT operation, which can be implemented very efficiently with a significant complexity reduction to $\mathcal{O}\left(NM\mathrm{log_{2}}\left(NM\right)\right)$, and the inverse of the diagonal matrix is much simpler than that of the full matrix in (\ref{eq_c_zf}) and (\ref{eq_c_mmse}). Besides, only the first column of the $NM\times NM$ channel matrix $\mathbf{H}$ is involved in our proposed equalizers, which further reduces the storage/memory cost.}
\end{remark}

\section{Computational Complexity Analysis}
This section analyses the detailed computational complexity between our proposed 2D FFT based linear equalizers and other ones in terms of the number of complex multiplications. As shown in Table \ref{tb}, the matrix inversion based ZF/MMSE equalizers \cite{hadani2017otfs} are most complex for the reason of full matrix inversion and multiplication operations, which both suffer a high computational complexity of $\mathcal{O}\left(\left(NM\right)^{3}\right)$. The FD-ZF equalizer proposed in \cite{ding2019otfs} utilizes the Kronecker product operation to avoid the full matrix inversion, however, it still suffers a computational complexity up to $\mathcal{O}\left(\left(NM\right)^{2}\right)$.

In this letter, the main operation of our proposed $\mathrm{FFT}_{2}$-ZF and $\mathrm{FFT}_{2}$-MMSE equalizers is the 2D FFT/2D IFFT operation with a computational complexity of $\mathcal{O}\left(NM\mathrm{log_{2}}\left(NM\right)\right)$. The analysis in Section \uppercase\expandafter{\romannumeral3} shows that the output of the $\mathrm{FFT}_{2}$-ZF equalizer includes three 2D FFT/2D IFFT operations and one diagonal matrix-vector multiplication. Similarly, the output of the proposed $\mathrm{FFT}_{2}$-MMSE equalizer includes three 2D FFT/2D IFFT operations, one diagonal matrix-vector multiplication, one diagonal matrix inversion and two diagonal matrix multiplications, and the computational complexity of the latter three operations is $\mathcal{O}\left(NM\right)$.

\section{Simulation Results}
This section shows the computational complexity and bit error rate comparison between the proposed 2D FFT based linear equalizes and other existing linear equalizers for OTFS. The simulation parameters are as follows: carrier frequency = $4$ GHz, subcarrier spacing = $15$ KHz \cite{raviteja2018interference}, the number of subcarriers $M$ = $512$, the number of OTFS symbols $N$ = $64$, the maximum speed = $200$ Kmph, and quadrature phase shift keying is used here. We use the 6-tap Typical Urban channel model \cite{channel1989} and the cyclic prefix duration = $5$ $\mathrm{\mu s}$.

\begin{figure}[H]
\subfigure[]{
\includegraphics[width=0.48154\hsize]{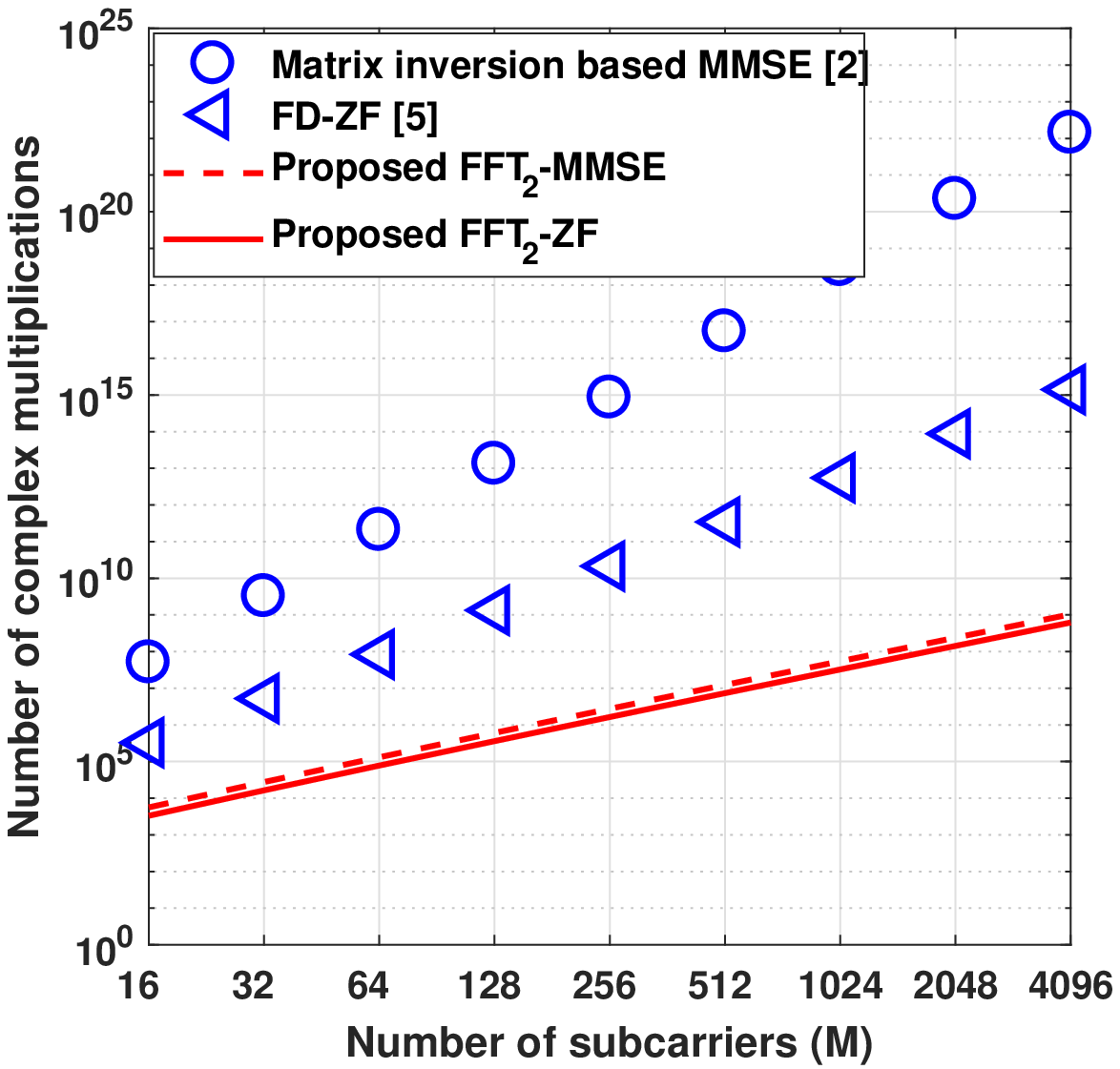}}
\subfigure[]{
\includegraphics[width=0.48154\hsize]{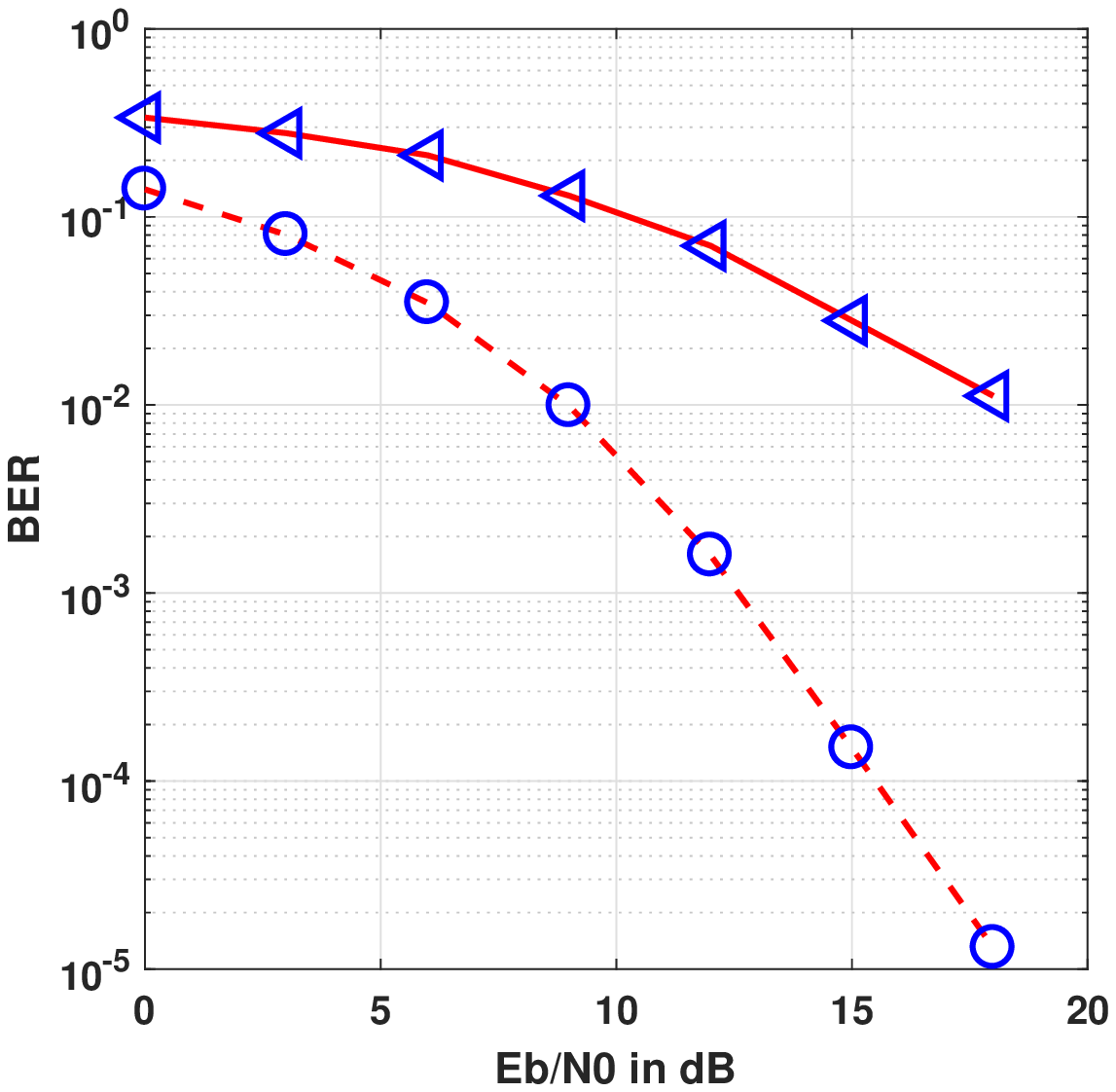}}
\caption{Performance comparison of different linear equalizers for OTFS. (a) Comparison of computational complexity; (b) Comparison of BER performance.}
\label{fg_sr}
\end{figure}

For the ease of complexity illustration, we assume $N$ equals to $M$ in Fig. \ref{fg_sr}(a), and the pilot-aided channel estimation scheme in \cite{raviteja2019embedded} is used here. From Fig. \ref{fg_sr}(a), it is obvious that our proposed $\mathrm{FFT}_{2}$-ZF and $\mathrm{FFT}_{2}$-MMSE equalizers have a much lower computational complexity than the matrix inversion based MMSE equalizer \cite{hadani2017otfs} and FD-ZF equalizer \cite{ding2019otfs}. Meanwhile, as shown in Fig. \ref{fg_sr}(b), the proposed $\mathrm{FFT}_{2}$-ZF and

%\begin{figure}[H]
%\centering
%\subfigure[]{
%\begin{minipage}{4.16cm}
%\centering
%\includegraphics[scale=0.373]{SR2_1.eps}
%\end{minipage}
%}
%\subfigure[]{
%\begin{minipage}{4.09cm} %距右侧的距离
%\centering
%\includegraphics[scale=0.373]{SR2_2.eps}
%\end{minipage}
%}
%\caption{Performance comparison of different linear equalizers for OTFS. (a) The comparison of computational complexity; (b) The comparison of BER.}
%\label{fg_sr}
%\end{figure}

\noindent $\mathrm{FFT}_{2}$-MMSE equalizers have the same BER as FD-ZF equalizer and matrix inversion based MMSE equalizer, respectively.

\section{Conclusions}
In this letter, we analyze the doubly block circulant feature of OTFS channel represented in the delay-Doppler domain, which inspires the proposed 2D FFT based ZF equalizer and MMSE equalizer. We demonstrate the computational complexity of the proposed linear equalizers is significantly reduced from $\mathcal{O}\left(\left(NM\right)^{3}\right)$ to $\mathcal{O}\left(NM\mathrm{log_{2}}\left(NM\right)\right)$. Simulation results further show that compared with other existing linear equalizers for OTFS, our proposed 2D FFT based ZF equalizer and MMSE equalizer enjoy a much lower computational complexity without any performance loss.

\bibliographystyle{IEEEtran}
\bibliography{ciations}

\end{document}